\documentclass[pre,aps,showpacs,amsmath,amssymb,twocolumn]{revtex4-1}

\usepackage{graphicx}
\usepackage{cancel}

\begin{document}


\title{Assisted crystal growing by tempering metastable vapor-liquid fluids}

\author{Gerardo Odriozola}  \email{godriozo@imp.mx}

\affiliation{Programa de Ingenier\'{\i}a Molecular, Instituto
Mexicano del Petr\'{o}leo, L\'{a}zaro C\'{a}rdenas 152, 07730
M\'{e}xico, D. F., M\'{e}xico}

\author{Felipe Jim\'enez-\'Angeles}  \email{fangeles@imp.mx}

\affiliation{Programa de Ingenier\'{\i}a Molecular, Instituto
Mexicano del Petr\'{o}leo, L\'{a}zaro C\'{a}rdenas 152, 07730
M\'{e}xico, D. F., M\'{e}xico}

\author{Pedro Orea}  \email{porea@imp.mx}

\affiliation{Programa de Ingenier\'{\i}a Molecular, Instituto
Mexicano del Petr\'{o}leo, L\'{a}zaro C\'{a}rdenas 152, 07730
M\'{e}xico, D. F., M\'{e}xico}

\date{\today}
\begin{abstract}
The metastable vapor-liquid coexistence of short-range attractive fluids hinders the formation of crystal nuclei, which in turn makes difficult the progress of the system towards its vapor-solid ground state. In this letter we show that crystal growth can be assisted by imposing temperature fluctuations. By so doing the obtained solid is nearly a fcc monocrystal in contrast with the extreme polycrystalline structure obtained at low temperatures. The study is carried out by combining the replica exchange Monte Carlo method and the standard slab technique. The obtained results suggest a pathway for growing coherent crystals from the metastable liquid. This is particularly relevant for the crystallization of globular proteins.
\end{abstract}


\maketitle

\section{Introduction}
\label{Intro}
The temperature-density phase diagram drastically changes with the range of the attractive part of the interparticle potential~\cite{Frenkel06,Benedek95,Sciortino05}. So that, when it is less than half of the particle's radius, the critical point locates below the freezing curve and the corresponding vapor-liquid phases are metastable~\cite{Chen08,Chen03,Lekker00,Lekker99,Frenkel07,Frenkel01,Prausnitz04,Lomakin96,Kumar05,Pagan05,Gast83}. This inhibits  crystal formation. To overcome this obstacle, it was proposed to take advantage of natural fluctuations occurring in the vicinity of the critical point, which may produce the first stable seed to grow the crystal phase~\cite{Frenkel97}. This pathway for crystallization was unsuccessfully applied to globular proteins solutions, which are known to follow the phase behavior described by short-range attractive model fluids~\cite{Frenkel97,Benedek95,Prausnitz04,Chen08}. Alternatively, temporal variations of protein concentration enhance crystallization, pointing out that fluctuations, though externally imposed, indeed promote nucleation and the crystal growth~\cite{Fraden07}.     

Computer simulations of short-range interacting fluids easily access and remain in the metastable region of the phase space, which signals a free energy barrier between the vapor-liquid and vapor-solid states~\cite{Kumar05,Rendon06,Lomakin96}. At sufficiently low temperatures, a liquid droplet evolves towards an imperfect solid made up of very small crystallites glued trough defective grain boundaries~\cite{Costa02}. On the other hand, it has been shown that crystallization is enhanced for the patchy square well potential fluid. In that case crystallization takes place after the formation of a high density droplet of self-assembled particles and well outside the metastable vapor-liquid region~\cite{Liu09}.

In this work we consider assisting the solid growth of an isotropic short-range interacting fluid by tempering its metastable vapor-liquid phases. In this way we expect the system to approach to the vapor-solid ground state. We also study the possibility of generating the solid seed at the vapor phase. To deal with these issues, we employed a simulation algorithm that imposes sudden temperature changes over several system replicas.

\begin{figure}
\resizebox{0.4\textwidth}{!}{\includegraphics{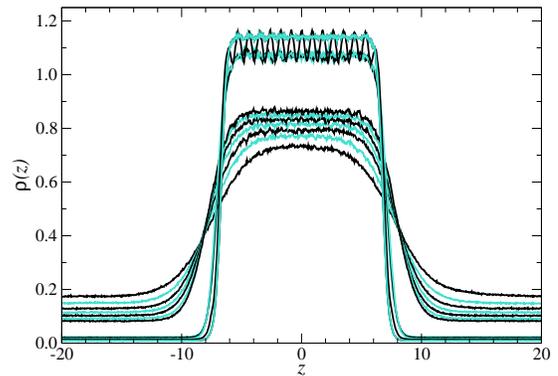}}
\caption{\label{profiles} Density profiles of the SW fluid for $\lambda = 1.15$ and for different temperatures. 
From bottom to top in the dense phase, temperature geometrically decreases from $T^*=0.562$ to $0.515$ by a factor of 0.992.}
\end{figure}

\section{Numerical Method}
\label{method}

Monte Carlo (MC) techniques are designed to sample form equilibrium. However, standard MC in some cases cannot drive the system to its ground state due to the existence of deep free-energy minima. In such cases, the replica exchange Monte Carlo (REMC) method frequently improves the sampling~\cite{Geyer91,Lyubartsev92,Marinari92,Hukushima96,Frenkel}. In this work we combine the REMC method with the slab technique~\cite{Frenkel}, i.e., a system initially containing a liquid slab surrounded by its vapor is replicated for different temperatures (parallel tempering). The general idea is to simulate $M$ replicas of the original system, being each replica at a different temperature, so that, the exchange of microstates among the cells is allowed (swap moves). These swap moves will make the replicas to follow some particular temperature routes, which can be then replicated by other simulation techniques. Thus, we use REMC to expose the replicas in the metastable state to temperature fluctuations. By so doing, we expect some of these replicas to escape from metastability.

Let us consider a model fluid made up of spherical particles at number density $\rho$, interacting through the attractive square-well (SW) potential, given by $u(r)=\infty$ for $r\le \sigma$, $u(r)=-\epsilon$ for $\sigma< r \le\lambda$, and $0$ for $r>\lambda$; being $\sigma$ the particles diameter, $\epsilon$ and $\lambda=1.15\sigma$ the well depth and width, respectively. For simplicity, 
we take the particles diameter as the unit length and $\epsilon$ is given in units of $k_BT$, being $k_B$ the Boltzmann 
constant and $T$ the absolute temperature. The corresponding reduced number density and temperature are given by $\rho^*=\rho\sigma^3$ and $T^*=k_BT/\epsilon$, respectively. 

We employed $M=12$ parallelepipeds with sides $L_x=L_y=8.0\sigma$ and $L_z=40.0\sigma$ for simulating the vapor-liquid 
interface of the SW fluid. Thus, each simulation box is initially set with all particles randomly placed within the slab, in such a 
way that the particles concentration is $\rho^* \approx$ 0.7 and the slab is initially surrounded by vacuum. This initial configuration is close to a metastable liquid-vapor state where liquid and vapor have similar volumes, which allows to search for seeding at both phases. In addition, we also set the bulk dense phase having a thickness larger than $10\sigma$. This is, in our experience, enough to avoid a shift in the phase diagram due to the presence of the interfaces between the dense and the vapor phases. Periodic boundary conditions are set in the three directions. The center of mass is placed at the box center. Verlet lists are implemented to improve performance. The highest temperature is set near but below the critical temperature. Other temperatures are fixed by following a geometrically decreasing trend. The swap acceptance rate is, in general, close to 20 \%. The replicas perform $10^{12}$ trials, while maximum displacements are varied to yield acceptance rates close to 30 \%. Long displacement trials are also considered. These displacements are important since they allow large jumps in the vapor phase with relatively large acceptance rates, while naturally carry out particle transference between both phases. After that, maximum displacements turn fixed and the thermodynamic properties are calculated over additional $10^{12}$ configurations. 

This study consists of computing the density profiles, radial distribution functions, average number of neighbors, and the order 
parameter $Q_6$, as a function of the temperature. The average number of first neighbors, $N_n$, is the average number of pairs having a center-center distance smaller than $1.2\sigma$ (the vectors
joining the centers of these pairs are named bonds). The order parameter $Q_6$
is defined as~\cite{Odriozola09,Steinhardt83,Rintoul96a}
$Q_6=\left(\frac{4\pi}{13}\sum_{m=-6}^{m=6}|<\!Y_{6m}(\theta,\phi)\!>|^2\right)^{1/2}$
where $<\!Y_{6m}(\theta, \phi)\!>$ is the average over all bonds
and configurations of the spherical harmonics of the angles $\theta$ and $\phi$ (these are the spherical angles of the
bonds measured with respect to any fixed coordinate system, since $Q_6$ is invariant). 

\begin{figure}
\resizebox{0.4\textwidth}{!}{\includegraphics{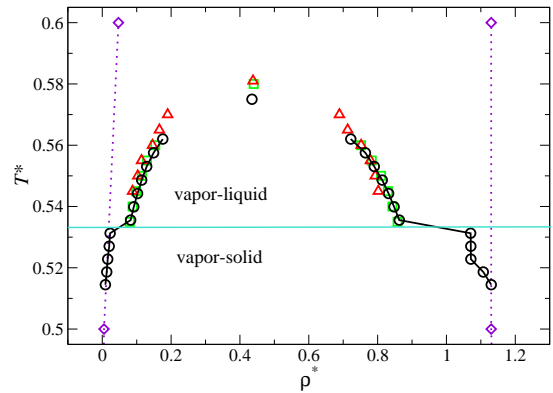}}
\caption{\label{phase-diagram} Phase diagram obtained from the condensed and vapor densities of Fig.~\ref{profiles} (open circles). Triangles and squares (red and green online) are metastable vapor-liquid data taken from ref.~\cite{Kumar05} and ref.~\cite{Rendon06}, respectively. Diamonds (purple online) correspond to the stable vapor-solid coexistence from ref.~\cite{Kumar05} (the dotted line is drown to guide the eye). The horizontal solid line divides the high and low temperature classes.}
\end{figure}


\section{Results and discussion}
\label{results}
From Fig.~\ref{profiles} it is observed that, as temperature decreases, at $T^*\approx 0.533$ the density profiles display an abrupt increase at the condensed phase while at the vapor phase a decrease occurs. The condensed phase of the fluid, with density $\rho^*_d$, is located at the box central region whereas the vapor phase, with density $\rho_v$, is contiguous to both sides of the condensed phase. The profile corresponding to the highest temperature ($T^*=0.562$) produces the lowest value of $\rho^*_d$ and the highest for $\rho^*_v$, whereas the one corresponding to the lowest temperature ($T^*=0.515$) yields the largest $\rho^*_d$ and smallest $\rho^*_v$. All the other density profiles are for intermediate temperatures. 

Thus, the density profiles appear divided in two classes: The high temperature class groups the density profiles with $\rho^*_d$ and $\rho^*_v$ in agreement with typical densities of liquid and vapor, respectively. This kind of density profiles corresponds to the metastable vapor-liquid coexistence. The low temperature class contains the higher values of $\rho^*_d$ and lower values of $\rho^*_v$. A distinction between the two classes of density profiles is the sharply defined interface for those at low temperature and the diffuse interface for those at high temperature. The latter is typical for a vapor-liquid coexistence, where the interface width increases with temperature. 

For the first simulation steps all density profiles behave like the high temperature class, since initial configurations are set in metastable states, i.~e., a liquid slab with density close to 0.7 and its corresponding vapor. At this stage, all swap trials have acceptance rates over 20\%. After a relatively large number of steps, driven by the temperature variations, a given replica escapes from its initial state towards its ground state, which give rise to the second kind of profiles. The breaking away from the metastable state is accompanied by a strong decrease of the configuration energy, which makes the REMC technique to preferable locate it at the lowest temperature. A similar process occurs for other replicas until the remaining high temperature replicas unlikely access temperatures below $T^* \equiv T^*_{s}=0.533 \pm 0.002$. In parallel, a free energy gap among the replicas at the metastable region and those closer to their ground state is developed. Due to this gap the swap moves between the high and low temperature replicas turn practically forbidden. Consequently, the high temperature replicas do not access low temperatures states and so, they cannot escape from the vapor-liquid metastable region. Under this rationale, not all replicas break away towards the ground state, but just some of them, implying that the whole system of replicas does not reach equilibrium. However, after approximately $3\times 10^{11}$ trials the whole system becomes stationary and thus, sampling is carried out. 

The well defined condensed-phase and vapor regions directly lead to the vapor and condensed densities, $\rho^*_v$ and $\rho^*_d$, by taking average at the different regions of the profiles. This is done without the need of fitting a hyperbolic function, as it is commonly used. Fig.~\ref{phase-diagram} shows the coexistence phase diagram obtained by plotting $\rho^*_v$ and $\rho^*_d$ as a function of temperature. The horizontal line points out the discontinuity of the $\rho^*_v$ and $\rho^*_d$ branches at $T^*\approx 0.533$. Such a discontinuity separates the two density profile classes discussed in Fig.~\ref{profiles}. 

For the upper part of the diagram we employed the exponential fitting (with $\beta=0.325$) to estimate the critical point~\cite{Frenkel}. We obtained $T^*_c=0.575$ and $\rho^*_c=0.435$ for the critical temperature and density, respectively. As temperature decreases (in Fig.~\ref{phase-diagram}), there is an abrupt shift from vapor-liquid to a denser phase-vapor coexistence. For $T^*<0.533$ we get the $\rho^*_d$ and $\rho^*_v$ results shown under the horizontal line. These branches correspond to the low temperature class concentration profiles in Fig.~\ref{profiles}. The density values of these branches indicate that the low temperature condensed phase is a solid. This means that the imposed temperature drops on the replicas aid the system to avoid the free energy barrier between the metastable and the ground state. That is, the crystal seeds are produced at low temperatures, which makes some of these microstates to appear at the other side of the free energy barrier when they are reset at high temperature in the swapping process.

Above the discontinuity, our results well agree with those previously reported and carried out without interfaces~\cite{Kumar05,Orkoulas10}. The results of ref.~\cite{Kumar05} are shown as triangles in Fig.~\ref{phase-diagram}. Our results also agree with previous calculations using the slab technique\cite{Rendon06}. Below the discontinuity, our data approach to the stable vapor-solid region described by Liu {\it et al.} (diamonds in Fig.~\ref{phase-diagram}). Actually, the vapor density branch falls over the interpolation of Liu's data, which were obtained by imposing a vapor-solid interface~\cite{Kumar05,Chen01}. In our case the vapor-solid interface is spontaneously formed. The observed agreement and the free energy decreasing confirms our assertion that above and below $T^*_{s}$ the data correspond to the metastable and stable coexistence diagrams, respectively.

\begin{figure}
\resizebox{0.4\textwidth}{!}{\includegraphics{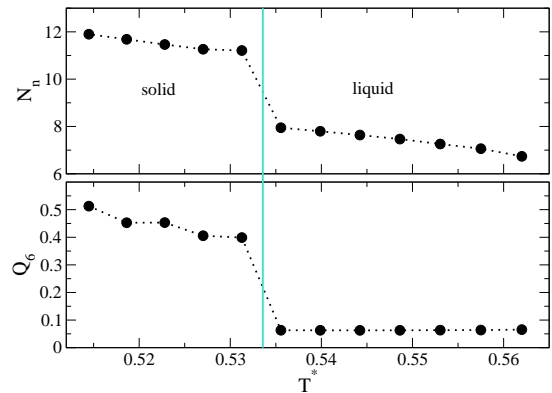}}
\caption{\label{Nn-q6} The first neighbor number, $N_n$, in the condensed region as a function of temperature (top). Order parameter $Q_6$ as a function of temperature (bottom). The vertical solid line corresponds to $T^*_{s}$.  }
\end{figure}

To characterize the high density phase, the first neighbor number, $N_n$, and order parameter $Q_6$ were followed as a function of temperature. The results are shown in Fig.~\ref{Nn-q6}, where a vertical line is plotted at $T^*_{s}$. These quantities were obtained only for the condensed phase, and avoiding the interfaces. For $T^*\ge T^*_{s}$, we obtained $N_n \approx 7$, and $Q_6 < 0.1$, which are the typical values for a liquid. At $T^*_{s}$, $N_n$ and $Q_6$ display a discontinuity: $N_n$ jumps from 7.95 to 11.20, whereas $Q_6$ from 0.064 to 0.400. It is known that the order parameter, $Q_6$, tends to $1/\sqrt{N_sN_n/2}$ for a completely random arrangement of particles, 0.5745 for a face cubic centered (fcc) crystal, 0.4848 for a hexagonal close packed (hcp) structure, and 0.3536 for a simple cubic geometry (being $N_s$ the number of pairs considered for the calculation). For $T^*<T^*_{s}$ we get $0.40 < Q_6 < 0.51$ and $11.2 < N_n < 12.0$, which signatures the presence of a solid condensed phase constituted by fcc or/and hcp crystal domains.

\begin{figure}
\resizebox{0.45\textwidth}{!}{\includegraphics{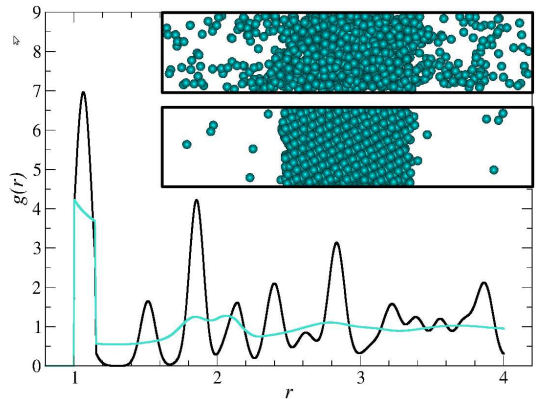}}
\caption{\label{Gdrs} Radial distribution functions (RDFs) for the condensed phases. The dark and light (blue) lines correspond to the lowest and highest computed temperatures, respectively. The insets are the corresponding snapshots for the vapor-liquid (top) and the vapor-solid (bottom) interfaces. }
\end{figure}

We gain further insight of the condensed phase structure by building the radial distribution functions (RDFs). This is shown in Fig.~\ref{Gdrs} for the lowest and highest computed temperatures. The high temperature RDF shows the typical behavior for a SW liquid, i.~e., there are two discontinuities at $r=\sigma$ and $r=\sigma+\lambda$, followed by two broad peaks at $r/\sigma \approx 1.85$ and $r/\sigma \approx 2.06$. At the location of the SW the RDF has large values, being the largest at contact, $r=\sigma$, and the lowest at $r=\sigma+\lambda$. There is a nearly linear decreasing trend at the SW region. On the other hand, the RDF for the lowest temperature shows several well defined peaks, pointing out a long range order. In this case, the highest peak is at $r/\sigma=1.06$, and the following are at $r/\sigma=1.51$, 1.85, 2.13, and 2.40, respectively, which are slightly shifted to the right with respect to those corresponding to a fcc crystal~\cite{Odriozola09}. This is in common for all obtained solid phases. This result contrasts with the rapid decay of the RDF for the extreme polycrystalline solid produced in the absence of temperature fluctuations, which has a density slightly above that of the liquid and a fractal structure~\cite{Costa02}. 

In practice, an equivalent procedure of the REMC simulation technique could be reached by successively imposing increasing and decreasing variations of temperature on a colloidal suspension. So that, in this way one could expect to produce a crystal seed at low temperature, which then can grow with the imposed fluctuating temperature. The growing process incorporates particles and new grains at the boundary of the growing nucleus~\cite{Costa02}. At this point, a large temperature aids reorienting the adjacent grains and relaxing their boundaries, however, a large exposure may lead to its total melting. On the other hand, a low temperature fastens the growing processes with the consequent incorporation of defects. Hence, tempering combines the reorienting, boundary relaxing, and growing processes of the nuclei, yielding a more coherent solid. Temperature fluctuations may also aid with subsequent grain coalescence processes.

Finally, we included two representative snapshots of the vapor-liquid (top), and vapor-solid (bottom) coexisting phases in the inset of Fig.~\ref{Gdrs}. In these snapshots many of the effects discussed throughout the paper are visualized. That is, the sharply defined vapor-solid interface (bottom), the fcc solid structure (bottom), the liquid structure (top), and a more dense vapor phase coexisting with the liquid than with the solid (top and bottom). The crystalline structure explains the presence and absence of oscillations in the vapor-solid density profiles of Fig.~\ref{profiles}: If any of the crystal planes is parallel to the interface ($xy$-plane), oscillations appear, whereas they vanish in any other case. This is why several low temperature profiles in Fig.~\ref{profiles} are flat even though they correspond to crystalline structures. The crystal growth cannot be controlled by the method, and so, the crystal planes may or may not be parallel to the interface. The snapshots also show the aggregation of vapor particles at both temperatures. However, no crystal nuclei are observed in the vapor phase in any case though the vapor is set relatively large in all cases.

\section{Conclusions}
\label{conclusions}

We studied the response of the SW fluid metastable liquid-vapor coexistence to temperature fluctuations under the scheme of the replica exchange Monte Carlo algorithm. Even though we imposed a large vapor volume, the solid seeding was observed always inside the liquid phase. This supports the hypothesis that the formation of a dense droplet is a necessary condition to produce the crystal~\cite{Liu09}. The obtained solid is in all cases a nearly fcc monocrystal in opposition to the highly polycrystalline structure obtained by a simple quenching process~\cite{Costa02}. Our results suggest that following the seed formation at low temperatures, the crystal phase can grow and relax approaching to the ground state when temperature is increased and decreased successively. 

\section{Aknowledgements}
The authors thank S. Fraden for a fruitful discussion of some key ideas presented here.


\end{document}